# TOWARDS SPECIAL DAEMON-SENSITIVE ELECTRON MULTIPLIER: POSITIVE OUTCOME OF MARCH 2009 EXPERIMENT

E.M. DROBYSHEVSKI and M.E. DROBYSHEVSKI

*Ioffe Physical-Technical Institute of RAS, 194021 St-Petersburg, Russia*
*E-mail: emdrob@mail.ioffe.ru*

Results of the experiments on daemon detection performed in St-Petersburg in March 2009 are presented. Adding the data obtained with the daemon-sensitive FEU-167-1 PM tubes to the data amassed in our previous measurements (starting from 2000) raises the confidence level of existence of the spring maximum in NEACHO (near-Earth almost circular heliocentric orbit) daemon flux to ~5$\sigma$.

The first test experiments conducted with the "dark" electron multiplier tubes, - TEU-167 with a thick (~0.5 µm) Al coating over all of the inner surface of the near-cathode multiplier section, including also its front screen, look encouraging. They provide supportive evidence for the existence of diurnal modulation of the daemon flux and offer ~$3.4 \times 10^{-7}$ cm$^{-2}$s$^{-1}$ for its lower limit in March, in good agreement with our earlier estimates and measurements.

## 1. Introduction. All experiments favor existence of daemons

Daemons (Dark Electric Matter Objects) are tentatively identified with the relic Planck objects (elementary black holes) with a mass of ~$10^{19}$ GeV which carry an electric charge of up to $Ze \approx 10e$ [1-4]. Negative daemons should be nuclear-active particles. In capturing a nucleus with an accompanying energy release of ~$10^2$ MeV, they initiate emission of atomic electrons, nucleons, and $\gamma$ rays, which makes them detectable [5]. We assumed also that while residing in the nuclear remainder (this complex was coined *c*-daemon), the daemon disintegrates in it nucleons, one after another, thus restoring the capacity of capturing a new nucleus. Finally, passage of the *c*-daemons possessing a varying electric charge through celestial bodies (the Sun, the Earth etc.) initiates in the course of celestial-mechanical evolution their capture from the galactic disk in the Solar System, first in strongly elongated, heliocentric, rosette-like orbits with perigees inside the Sun. Most of these objects enter eventually the interior of the Sun to form its daemon kernel. Gravitational interaction of some of the daemons with the Earth transfers (and accumulates) them in strongly elongated Earth-crossing heliocentric orbits (SEECHOs) and, subsequently, in near-Earth almost circular heliocentric orbits (NEACHOs) and geocentric Earth-surface crossing orbits (GESCOs; the latter build up in the Earth to produce a daemon kernel, a process accounting for a variety of geophysical phenomena). Ground-level fluxes from the three latter types of orbits exceed the flux of galactic daemons by 4-5 orders of magnitude, with the SEECHO daemon flux varying with an annual period (with the maximum in early June), and the NEACHO and GESCO objects, with a half-year period (with maxima in March and September) [6].

The daemon paradigm is supported presently not only by our experiments (at a confidence level of ~5$\sigma$, see Sec. 5 below) using thin ZnS(Ag) scintillators [5] and DAMA experiments (at a level of >8$\sigma$) [7,8] based on large assemblies (of up to 250 kg) of NaI(Tl) crystals but by the lack (fully understandable [9]) of sensible results from the other detectors on the lookout for other candidates for DM (either the sensitivity of these detectors is too low for them to detect daemons or, conversely, the data treatment techniques employed preclude recording of very strong daemon-associated events).

In the course of the experiments, we have revealed a variety of aspects of daemon interaction with matter which, while originally having been unexpected, were later found to provide supportive evidence for the basic working hypothesis, thus refining it. In 2005 we found [10], in particular, that some anomalies find reasonable explanation if we assume that some PM tubes themselves respond to daemons crossing them (before this finding we did not distinguish PMTs).

We are reporting here on further development and application of this observation. We are describing the results of the experiments performed in March 2009, both with the FEU-167-1 tubes found earlier to be daemon-sensitive and with specially designed electron multipliers [11] based on FEU-167 (unlike their original version, which was light sensitive, these TEU-167 are "dark" electron multipliers - DEMs), which we believe to become, after some modifications, efficient detectors of daemons which initiate electron emission in interaction with matter.

**2. Possible mechanism of the PM tube response to daemon passage**

Excitation of a pulse in the passage of a daemon through a PM tube was corroborated by Baksan underground experiments with a simplified version of our detector [12]. The triggering signal in its first oscilloscopic trace was generated by a long (leading pulse edge 2-3 µs) heavy particle-type scintillation (HPS) in the ~10-µm-thick ZnS(Ag) layer, while the noise-like signal (NLS - with a fast leading edge of ~1-1.5 µs), shifted in time by $\Delta t$ relative the HPS, was initiated by the daemon crossing a light-insulated FEU-167-1 which was selected for its highest sensitivity to such an event.

Our analysis of the possible factors which could activate a PMT to a daemon crossing it ended up with the simplest assumption [10] that in passing through vacuum in the PM tube the $c$-daemon can not only conserve its negative charge $|Z_{eff}| \geq 1$ but even increase it, because it does not capture a new nucleus in vacuum. Therefore, in passing through the PMT wall, such an object is capable of capturing a nucleus, accompanied by electron emission, already in moving a distance ≤1 µm. Actually, this is a process similar to excitation by a negative $c$-daemon of an HPS in ZnS(Ag) [5]. These electrons generate in the PMT a detectable electric pulse, in the same way as this occurs with photoelectrons. Significantly, the number of emitted electrons should depend here on the possibility of capture and reemission by the excited nucleus of additional, refilling electrons from the surrounding matter. Obviously enough, the latter process is most efficient in metals with their collectivized electrons.

Indeed, induction measurements revealed [10] that some of the FEU-167-1 multipliers (Fig. 1) most sensitive to daemon passage have, by and large, a noticeably thicker (up to ~0.5 µm) Al mirror coating on the internal (cylindrical side and conical rear) surfaces of the cathode section of their glass bulb, for the standard average thickness of ~0.1 µm. The Al coating provides electric contact with the photosensitive antimony-alkali metal film (which is deposited on all of the surface of the cathode part of the bulb, both on the front disc and the Al mirror) and can partially prevent light from entering the near-cathode volume from aside and from the rear (which reduces also the CR background, see below Sec. 5). The thickness of the "mirror" coating does not affect directly the photometric properties of the PMT.

To compare the results, we conducted in March 2009 experiments with the old, daemon-sensitive bottom FEU-167-1 (this is No00099 in the former module 3 [10] used also in Baksan [12] in 2005-2006 (presently this is module 23), and No00068 in the former module 4 (presently module 3), see Table 1), as well as with two modified TEU-167; the latter differ from the FEU-167 tubes in a continuous (including the front disc also) internal, fairly thick (up to ~0.5 μm) Al coating (the bottom instruments in modules 1 and 4).

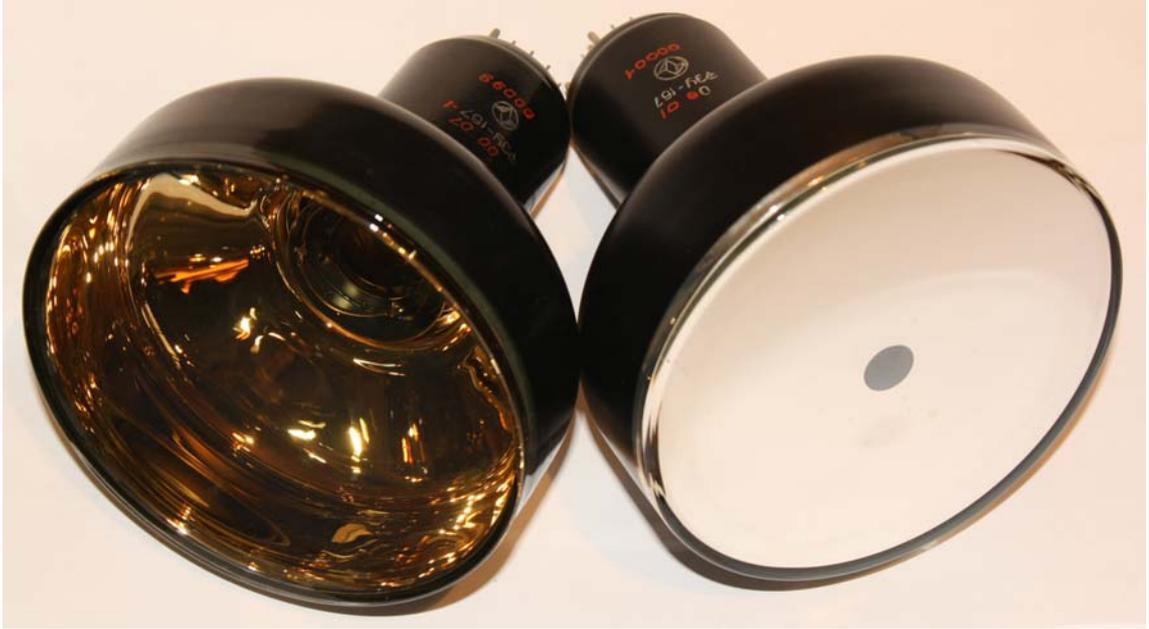

Figure 1. PM tube of FEU-167-1 type (its front disc – photocathode – has ø125 mm), and DEM tube of new TEU-167 type [11] with the front disc also covered on the inside with ~0.5 μm Al layer; it has ø13 mm central transparent window for the light calibration of the instrument.

Table 1. March $N(\Delta t)$ distributions (-100 < $\Delta t$ < 100 μs) obtained in 2000-2005 [10], 2006 (Baksan [12]), and in 2009 (this paper, see text). Ten bins, each 20 μs wide, are centered at the $\Delta t$ specified in the Table. Σ is the total number of the time-shifted ($|\Delta t|$ > 0.4 μs) double events with HPS in the first (triggering) channel; m is the number of modules used in detector; Σ/mT is the rate of event records per hour per module (the real life times are somewhat smaller than T's given).

| Dates of exposition, its duration T (hrs), number of a module | | $N(\Delta t)$ | | | | | | | | | | Σ | m | Rate of records Σ/mT |
|---|---|---|---|---|---|---|---|---|---|---|---|---|---|---|
| | | -90 | -70 | -50 | -30 | -10 | +10 | +30 | +50 | +70 | +90 | | | |
| 24.02-27.03.2000 T = 700h | | 24 | 19 | 27 | 11 | 25 | 19 | 39 | 15 | 18 | 15 | 212 | 4 | 0.076 |
| 24.02-27.03.2003 T = 676h | | 25 | 29 | 31 | 23 | 31 | 22 | 38 | 33 | 30 | 23 | 285 | 4 | 0.105 |
| 08.03-20.03.2004 T = 288h | | 3 | 6 | 7 | 7 | 7 | 3 | 10 | 5 | 4 | 10 | 62 | 2 | 0.108 |
| 05.03-25.03.2005 T = 480h | 1, 2, 24 | 14 | 17 | 15 | 18 | 20 | 20 | 25 | 15 | 15 | 18 | 177 | 3 | 0.123 |
| 05.03-25.03.2005 T = 480h | 3 | 40 | 52 | 55 | 40 | 32 | 51 | 68 | 54 | 64 | 41 | 497 | 1 | 1.035 |
| | 4 | 67 | 63 | 62 | 68 | 64 | 53 | 70 | 56 | 56 | 54 | 613 | 1 | 1.277 |
| | 23 | 19 | 14 | 21 | 25 | 26 | 20 | 29 | 18 | 36 | 22 | 230 | 1 | 0.479 |
| 03-18.03.2006 T = 348h | 3 | 15 | 17 | 17 | 11 | 10 | 24 | 23 | 18 | 15 | 11 | 161 | 1 | 0.462 |
| 18-24.03.2006 T = 164h | 3 | 4 | 10 | 12 | 14 | 7 | 9 | 13 | 8 | 9 | 8 | 94 | 1 | 0.573 |
| 08-22.03.2009 T = 336h | 3 | 12 | 14 | 15 | 12 | 16 | 13 | 19 | 19 | 8 | 12 | 140 | 1 | 0.417 |
| | 23 | 5 | 5 | 7 | 10 | 10 | 8 | 12 | 11 | 11 | 4 | 83 | 1 | 0.247 |

## 3. Experiment with daemon-sensitive FEU-167-1 PMTs

In trying to reduce the background, we started the experiment on February 24, 2009 with a module-by-module replacement of the B3-s-type ZnS(Ag) phosphor (manufactured in 1972) with an FS-4 scintillator acquired specially for the purpose (with the same powder dispersion). We have not been able to reveal any differences in their operation.

The scintillation screens were fabricated this time from transparent, 3-mm-thick polystyrene in the form of 4-cm-high, $50 \times 50$ cm$^2$ boxes, with a 3.5 mg/cm$^2$ ZnS(Ag) powder layer (FS-4 scintillator) deposited on the *upper* surface of the box bottom. The box was covered on top with a lid, likewise of polystyrene. The distance from the ZnS(Ag) layer to the screen of the top (triggering) FEU-167 was 22 cm, and that to the screen of the bottom PMT covered with aluminum-coated Lavsan film was 29 cm. Just as before [5,10,13], the scintillation screen was mounted inside a tinned-iron sheet cubic box, 51 cm on a side, whose top lid (intended to enhance on purpose the detector asymmetry with respect to the up/down direction [5]) was made of two sheets of black paper. To reduce the number of signals from these 'noisy' FEU-167-1 (in the period from 24 February to 8 March, 2009 (285 h altogether), each of these PMTs produced about 970 double events) whose face screens were made from K-containing C-52-1 glass (SiO$_2$ - 68.7±1.2%, B$_2$O$_3$ - $19^{+0.5}_{-1.0}$%, Na$_2$O - 4.4±0.4%, Al$_2$O$_3$ - 3.5±0.5%, K$_2$O - 4.4±0.5%), we reduced on 8 March 2009 the supply voltage of both the top and the bottom PMTs in all (four, see also below) modules by 70 V (note that a change of the FEU-167 voltage by 100 V changes the sensitivity by close to one half). Starting from March 8, 2009, the modules were blown by nitrogen vapor flow (gas flow rate ≈ 12 cm$^3$s$^{-1}$ per module).

The results of the present experiment are summed up in the two last lines of Table 1.

We readily see that the +30 µs-maximum in the $N(\Delta t)$ distribution is not very pronounced, exactly as this should be expected if we recall the drop in the sensitivity of the system caused by the decrease of the power supply voltage.

Nevertheless, if we just add all the $N(\Delta t)$ March distributions for the years 2000, 2003, 2004, 2005 (St-Petersburg [10]), 2006 (Baksan [12]), and 2009, we find that the +30 µs-maximum contains 346 events for a total number of double shifted events of 2554 recorded in the interval $\Delta t = \pm 100$ µs, which were triggered by HPSs from the top PMTs[a] (we do not consider here, as always, the "unshifted" events with $|\Delta t| \leq 0.4$ µs, with the vast majority of them being excited by cosmic rays). This translates readily to a confidence level (C.L.) of 4.87$\sigma$ for the +30 µs maximum caused by the NEACHO daemon flux.

## 4. Notes on possible reasons for the incomplete reproducibility of results

Reproducibility of the results of an experiment is, on the one hand, a proof of their being correct, while on the other it provides arguments for the validity of the hypotheses underpinning this experiment. Interestingly, the practically unavoidable deviations from the expected results are also of considerable significance. These deviations permit one to refine the starting concepts and models.

We encountered a lack of reproducibility of our first positive results of March 2000 already in April-May 2000. Only one year later did we understand [13] that it originated from the half-year variation in the daemon flux, which provided an insight [6,10,12,13] into the key part played by celestial mechanics in the creation, evolution, and buildup of the daemon population in near-Earth orbits, in particular, in NEACHOs, from where they fall on the Earth with $V \approx$ 10-15 km/s, as this followed already from the first measurements [5]. The poor results obtained in another laboratory room, smaller in area but air-conditioned (and with the windows closed tightly) revealed the role of radon in an increase of the background level [12]. The differences observed in the behavior of detector moduli using FEU-167 multipliers of different Manufacturers focused our attention [10] on the "personalities", so to say, of these instruments originating from subtle aspects of interaction of daemons with matter, some of which still remain unclear. Actually, it is the latter observation that has initiated the present study.

---

[a] To avoid ambiguities, we have to note the $N(\Delta t)$ given in [5] (Fig. 2) contains some events with HPSs in the second trace, while $N(\Delta t)$ in [13] (Fig. 1) contains also events with two signals (one of them is HPS, another – NLS) in the first trace.

Irreproducibility of results, however, is far from always caused by our lack of understanding of the properties of the objects under investigation. For instance, the initial conditions may change for reasons out of control of the experimenter.

Indeed, as seen from Table 1, the average rate of detection of events shifted in time by $|\Delta t| > 0.4$ µs within an interval $\Delta t = \pm 100$ µs, with a single HPS on the first (triggering) channel and fairly weak (≤1.6 mV) NLS on the second, was in March 2000 about 0.076 events/hour/module. The C.L. of the +30 µs maximum in the $N(\Delta t)$ distribution was $2.85\sigma$ (99.56%) (here and subsequently, we are presenting excess digits and some details of the calculations solely for the conveniency of the Reader). Five years later, the rate of detection increased to 0.123 events/hour/module (see Table 1). With such a rate of detection, the significance of the peak of interest to us would have dropped down to $2.28\sigma$ (97.74%), so that it would not have attracted our attention.

Now what could be possible reasons for the rise in the number of detected events (primarily naturally background signals)? Two fairly obvious ones deserve mentioning.

It is known that the flux of high-energy galactic cosmic rays which create atmospheric showers (and some NLSs in our case) varies in opposite phase with the solar cycle [14,15]. In 2000, the Wolf number were approaching their maximum. The maximum persisted for about three years, but today (2009) we are in the region of the minimum where the radiation background generated by cosmic rays reaches its highest level.

On the other hand, the strong increase of electromagnetic cross-talk in our case is much more impressive. Any experimentalist is aware of the fact that success in suppressing interference depends largely on the skill of the researcher. Indeed, if the room is not screened thoroughly, even a slight change in cable position may sometimes affect the background level.

What accounts for the drastic rise in electromagnetic background observed in the recent years? The Ioffe Physical-Technical Institute of the Russian Academy of Sciences (St-Petersburg), the largest in Russia, is surrounded by maybe ten other Research Institutes within a half-kilometer radius. After the collapse of the USSR, in 2000 the work at these Centers and at our Institute practically ceased. The situation has changed in the recent years, however, scientists started to work again, largely with instruments radiating high-level noise which sometimes is difficult to discriminate from useful signals (this relates particularly to small-amplitude NLSs) and can initiate continuous triggering of the system, with its memory overfilling after a few hours of operation while it is intended for months of operation (to site just one example, these uninterruptible power supplies in wide use presently emit in radio range).

It thus appears that we have been lucky by having started our pursuit of daemons in late 1990s, with the lowest background level of interference of both types.

## 5. Experiment with modified PMTs (now these are DEMs, for "dark electron multipliers")

The reasoning of Sec. 2 suggested a conclusion [10,11] that in order to enhance the sensitivity of a PMT to daemons crossing it one should increase the thickness and/or change the composition of the coating, as well as to extend this coating to the front disc too, thus making the PMT "blind" (leaving only its capacity of multiplying the electrons created in the interaction with material of daemons rather than light). On our order, such FEU-167-based DEMs (now TEU-167) were manufactured by Screen-Optical Systems Company in Novosibirsk (Fig. 1). The thickness of the aluminum coating of the cathode section, including also the screen, was as large as ~0.5 μm (because of the specific features inherent in the technology of Al mirror deposition by vacuum evaporation, the thickness could be nonuniform, and at the "corners" of the section, for instance, in the regions where the disc joins the cylinder, it could be less ~ by one half). A dia. 13-mm transparent window was left unobscured at the screen center to allow calibration with light. All of the inner surface of the cathode section was, as usually, coated by the photosensitive Sb-K-Na-Cs composition.

Two standard (see Sec. 3) moduli 1 and 4 equipped with such TEU-167's as bottom sensitive elements (their screens were covered by aluminum-coated Lavsan film) were used in an experiment in St-Petersburg which lasted from February 24 to March 03, 2009. Because the number of double shifted events registered with these modules in the period from February 24 to March 08, 2009 (in ~285 h) also turned out unexpectedly large (775 in module 1 and 919 in module 4), the voltage applied to their top and bottom multipliers was reduced, as in modules 3 and 23 (see Sec. 4), by 70 V (we are going to consider in what follows primarily the results amassed from March 08 to March 22, 2009; in 336 hours, module 1 recorded 1063, and module 4, 773 events; the practical absence of the effect of reduced voltage on the rate of event recording may indicate a substantial contribution of electromagnetic cross-talk background).

In the course of these experiments, a sharp drop in the count rate of unshifted ($|\Delta t| \leq 0.4$ μs) NLSs generated by cosmic ray showers was observed. An analysis of this observation led us to the conclusion [16] that the shower particles are detected due to their exciting light pulses in the glass itself of the DEM screen, in this particular case - near its dia. 13-mm transparent hole rather than in the Al or light-sensitive Sb-K-Na-Cs coatings or in the dynode assembly of the electron multiplier.

We tried two ways in an attempt at detecting any indication of the March maximum in the NEACHO daemon flux.

First, while knowing that the primary NEACHO flux (downward in the Northern hemisphere) lasts as low as ~10 days at most [6,12], we analyzed only large $U_2$-amplitude NLSs by the second trace for the period from February 24 to March 08, 2009 to select double events with a total number corresponding to ~1 event/hour/module. We obtained 337 events with $U_2 \geq 1.2$ mV for module 1 ($N(\Delta t)$ for the ±100-μs interval is 31, 30, 35, 35, 27, 37, 38, 38, 29, 37) and 297 events with $U_2 \geq 1.0$ mV for module 4 ($N(\Delta t)$ is 28, 28, 29, 29, 30, 34, 36, 26, 26, 31). While one does indeed discern a maximum in the +30-μs bin, its significance is only 1.23$\sigma$ (but if we add these distributions to the data of Table 1 in Sec. 3, its C.L. becomes **4.94$\sigma$**!). Application of similar procedure to the interval from March 08 to March 22, 2009, in which a sizable part of the flux with $V \approx 10$ km/s is made up of daemons captured into GESCOs and propagating both down- and upward, did not end up with reasonable results. This finds a ready explanation; indeed, our DEMs should respond in a nearly the same way to both downward and upward propagating fluxes, and the appearance in mid-March in the $N(\Delta t)$ distribution, besides the +30 μs, of a -30 μs maximum too, and, moreover, their shift with time toward larger $|\Delta t|$ as a result of GESCO daemon deceleration by the Earth's material [13], could only reduce the significance of both maxima.

Second, to discriminate between the up- and downward fluxes, it would certainly be desirable to locate in the detector a factor (parameter) depending on the direction of daemon propagation.

**6. The detector reaction asymmetry favors the daemon paradigm once again**

Our system has such a parameter. It is the area (width) of the HPS pulse normalized to its amplitude. As far back as 2003, we discovered [13] that this area $S$ depends on the direction in which a daemon crosses the ZnS(Ag) layer deposited on polystyrene. If a daemon, which has captured, say, a Zn nucleus in the scintillator, moves into the bulk of the polystyrene, part of the particles emitted by this $c$-daemon will stop in it and will not reach the ZnS(Ag) layer. The statistical mean of the length of the scintillation will be less than if the c-daemon had moved in the opposite direction - into vacuum or air - where nothing would prevent the particles emitted by it from reaching the ZnS(Ag) layer.

Therefore, if we split $N(\Delta t)$, as we did it in [13], into two parts, one of which, $N_w(\Delta t)$, would contain events with "wide" HPSs in the $S > S_m$ wing, and the other, $N_n(\Delta t)$, those with "narrow" HPSs ($S < S_m$, where $S_m$ is the mean of the distribution, which is about a Gaussian for the HPSs, see Fig. 2 in [13]), then for an upward daemon flux (i.e., in this our case, where the ZnS(Ag) layer covers the top surface of the polystyrene) the left-hand ($\Delta t < -0.4$ μs) part of $N_w(\Delta t)$ will contain under ideal conditions, i.e., where all the upward-moving daemons produce HPSs with $S > S_m$, an excess of events, which will reflect the daemon flux. And conversely, the downward-moving daemon flux can be identified with an excess of events in the right-hand part ($\Delta t > 0.4$ μs) $N_n(\Delta t)$ distribution. Obviously enough, if there are fluxes going in both directions, adding $N_w(\Delta t)$ to its "mirror" distribution $N_n(-\Delta t)$ will produce an excess of events in the left-hand side of the sum of the distributions, $N_w(\Delta t) + N_n(-\Delta t)$, compared with its right-hand part.

It is obvious that when there are no fluxes, and the signals produced are actually the background only, the $N(\Delta t)$ and $N_w(\Delta t) + N_n(-\Delta t)$ distributions should be statistically indistinguishable.

We have performed the above procedure with the data from modules 1 and 4 for the period of March 08 to March 22, 2009 (~336 hours), and rejected 923 "extreme" events initiated by NLSs with a small ($U_2 \leq 0.6$ mV) and a large ($U_2 \geq 1.2$ mV) amplitudes, which are most likely to have been triggered by strays. Of the 913 events left, with $0.6 < U_2 < 1.2$ mV, 478 events fell into the left-hand ($\Delta t < -0.4$ μs) wing of the $N_w(\Delta t) + N_n(-\Delta t)$ distribution, and 435 events, into the right-hand one. While the statistical significance of this difference is not very high (a cautious $\chi^2$-criterion estimate yields C.L. ≈ 84% for this difference), it signals the existence of an upward daemon flux $f \approx (478 - 435)/(2 \times 115 \times 336 \times 3600) = 1.5 \times 10^{-7}$ cm$^{-2}$s$^{-1}$ (and of the same flux downward, if they are equal) (here 115 cm$^2$ is the DEM screen area, and 336 h is the exposure time). The value thus obtained matches satisfactorily with our earlier measurements of the flux [10,12] and the DAMA data [7,8] treated in the context of the daemon approach [9].

It is possible to amplify the effect by taking into account one more factor contributing to the asymmetry of the daemon detection process. Indeed, daemons catch up with the Earth by striking it from the outermost NEACHOs which, on the average, touch the Earth's orbit sometime a week or so before the equinox [6,12]. In March, the maximum of the downward NEACHO daemon flux in our Northern hemisphere falls on evening hours, as pointed out before [12]. Therefore, if we apply the above procedure to the data obtained from March 08 to March 22, 2009 within the interval from 15 to 23 h, we find that the left-hand wing of the $N_w(\Delta t) + N_n(-\Delta t)$ distribution contains 175 events, and the right-hand one, 143 events. Despite the small number of the events, the significance of this difference increases with respect to the value quoted in the preceding paragraph to C.L. ≈ 93%, and the detected flux of daemons rises to $f \approx 3.45 \times 10^{-7}$ cm$^{-2}$s$^{-1}$. Obviously enough, this value should be considered as the lower estimate only, if for no other reason that $S$ lies on the same side of $S_m$ not for all the HPS events triggered by the upward (or, conversely, downward) moving daemons. Moreover, the efficiency of the detector itself is far from being 100%. (We note that for the night/morning hours ($23^h$-$7^h$) the left-hand wing of $N_w(\Delta t) + N_n(-\Delta t)$ contains 163, and the right-hand one, 148 events, which yields $f \approx 1.60 \times 10^{-7}$ cm$^{-2}$s$^{-1}$ with the C.L. $\approx$ 76% only).

If the flux does indeed increase in the evening hours, this may indicate that the fall of NEACHO daemons continues to play a significant role in mid-March as well.

## 7. Conclusion: The potential of the daemon paradigm and the immediate tasks to be undertaken

As we see, the daemon-based concept of the nature of Dark Matter is supported convincingly not only by direct experiments aimed at detection of these Planck objects (or of their non-detection if experiments are aimed at a search for various WIMPs) but turns out to be fruitful, both in providing reasonable interpretation for many observations and revealing intriguing aspects of the DM problem and suggesting new (sometimes unexpected) ways to observation of the daemons themselves, which are inspired by these aspects.

The goal of our work as we see it lies not just in routinely accumulating statistics of observations (this is rather a by-product of our study) but in gaining insight into the essence of interaction of daemons with matter, including the celestial mechanics underlying their evolution.

This study has once more substantiated the possibility of direct detection of daemons with FEU-167-1 PMTs with an enhanced thickness of the inner Al coating. Adding our new measurements (see Sec. 5) to the data accumulated in the recent years has raised the confidence level of existence of the March maximum in the flux from NEACHOs to $5\sigma$.

The main result of the March 2009 experiments discussed here lies, however, in demonstrating that purposefully modified FEU-167s (referred to presently as TEU-167s, in which not only the rear and the sides of the near-cathode section of the bulb but the screen itself also are coated on the inside by a thick Al layer) do indeed detect the passage of daemons with $V \approx$ 10-15 km/s. Although these first experiments have not thus far produced high-confidence-level results (the statistics are not yet good enough for that), they speak for themselves. Considered in the context of the basic ideas underlying the interaction of daemons with matter developed earlier (take, for example, a somewhat unexpected dependence of scintillation width on the direction of daemon passage through the ZnS(Ag) surface layer), these data provide information on the daemon flux, which correlates with our earlier measurements and the DAMA observations,[b] if the latter are treated in the frame of the daemon paradigm. Moreover, there are indications that this flux undergoes the diurnal variation of the type predicted before [12], and that it also favors (although we still have not realized how one could translate all these "Favors" numerically to a rigorous C.L. estimate) our celestial-mechanics-based scenario of the daemon capture into SEECHOs and, subsequently, into NEACHOs and GESCOs.

Our nearest task, as we see it, lies in refining the method of daemon detection developed by us with the use of properly modified DEM tubes. In view of the important part played by the daemon-sensitive PMTs, it appears appropriate to suggest that our first experiments [5,13] with two separate thin ZnS(Ag) scintillators detected daemons not through operation of the bottom ZnS(Ag) layer (it gave no observable HPSs) but rather due to the bottom FEU-167's having been sensitive somewhat to the passage of daemons which did excited the NLSs in them. Making another step along these lines, we could suggest that some researchers elsewhere did try to reproduce our first experiments but failed in obtaining unambiguous results because they used PMTs of different types.

---

[b] It is instructive to note that, besides the first DAMA/LIBRA results [8], the latest ones [17] do also confirm the daemon-paradigm-based predictions [9] made two years ago.